\documentclass[10pt]{article}
\usepackage{amsmath,amsfonts,latexsym, amssymb}
\usepackage{epsfig}

\usepackage{color}

\newcommand{\wt}{\widetilde}
\newcommand{\wh}{\widehat}

\newcommand{\e}{\varepsilon}
\newcommand{\eps}{\varepsilon}
\newcommand{\pt}{\partial}
\newcommand{\rd}{{\rm d}}
\newcommand{\bR}{{\mathbb R}}

\newcommand{\bb}{{\bf{b}}}

\newcommand{\al}{\alpha}

\newcommand{\be}{\begin{equation}}
\newcommand{\ee}{\end{equation}}

\newcommand{\om}{{\omega}}

\newcommand{\cL}{{\mathcal L}}

\newcommand{\cY}{{\mathcal Y}}

\newcommand{\cK}{{\mathcal K}}

\newcommand{\E}{{\mathbb E }}
\newcommand{\R}{{\mathbb R }}

\renewcommand{\P}{{\mathbb P}}
\newcommand{\bC}{{\mathbb C}}

\newtheorem{theorem}{Theorem}

\newtheorem{lemma}[theorem]{Lemma}
\newtheorem{proposition}{Proposition}

\newtheorem{remark}{Remark}

\newcommand{\qed}{\hfill\fbox{}\par\vspace{0.3mm}}

\numberwithin{equation}{section}
\numberwithin{theorem}{section}
\numberwithin{definition}{section}
\numberwithin{proposition}{section}
\numberwithin{remark}{section}

\usepackage{geometry}     
\usepackage{graphicx}

\numberwithin{equation}{section}
\def\cal{}
\def\RR{{\mathbb R}}

\def\EE{{\mathbb E}}\def\PP{{\mathbb P}}
\def\NN{{\mathbb N}}

\def\W2{W^{1,2}({\cal O}(M))}

\def\1half{\frac{1}{2}}

\input epsf
\begin{document}

\title{Bulk Universality  for Wigner Matrices}

\author{L\'aszl\'o Erd\H os${}^1$\thanks{Partially supported
by SFB-TR 12 Grant of the German Research Council}, 
Sandrine P{\'e}ch{\'e}${}^2$, Jos\'e A. 
Ram{\'\i}rez${}^3$, \\
Benjamin Schlein${}^4$,\;
and Horng-Tzer Yau${}^5$\thanks{Partially supported
by NSF grants  DMS-0757425, DMS-0804279} \\
\\ Institute of Mathematics, University of Munich, \\
Theresienstr. 39, D-80333 Munich, Germany${}^1$ \\ \\
Institut Fourier, University of Grenoble 1,\\
100 rue des Maths, BP 74, 38402 St. Martin d' H{\`e}res, France${}^2$ \\ \\
Department of Mathematics, Universidad de Costa Rica\\
San Jose 2060, Costa Rica${}^3$ \\ \\
Department of Pure Mathematics and Mathematical Statistics
\\  University of Cambridge \\
Wilberforce Rd, Cambridge CB3 0WB, UK${}^4$ \\ \\
Department of Mathematics, Harvard University\\
Cambridge MA 02138, USA${}^5$ \\ \\
\\}

\date{Oct 20, 2009}

\maketitle

\begin{abstract}

We consider  $N\times N$ Hermitian Wigner random matrices 
$H$ where the probability density
for each matrix element is given by the density $\nu(x)= e^{- U(x)}$. 
We prove that the eigenvalue statistics in the bulk is given by  
Dyson sine kernel provided that $U \in C^6(\RR)$ 
with at most polynomially growing derivatives and
$\nu(x) \le C\, e^{ - C |x|}$ for $x$ large. 
The proof is based upon an approximate time reversal of 
the Dyson Brownian motion combined with the convergence of
the eigenvalue density to the Wigner semicircle law on
short scales.

\end{abstract}


\medskip

{\it Running title:} Universality for Wigner matrices

\medskip



\bigskip
\section{Introduction} 

The fundamental reason why  
random matrices have been used to model many large systems 
is based on the belief  that their local eigenvalue statistics 
are universal. This is generally referred to as the universality 
of random matrices. 
It is well-known that the local behavior of eigenvalues near the spectral edge
and in the bulk are  governed by the Tracy-Widom law and by the
Dyson sine kernel, respectively.
Since  the seminal work of Dyson  \cite{Dy1} 
for the Gaussian Unitary Ensemble (GUE), the  universality both for the edge
 and the bulk were proven for very general classes of unitary
 invariant ensembles
in the past two decades (see, e.g. \cite{M, PS, D,BI,DKMVZ1,DKMVZ2, LL} 
and references therein).
 For  non-unitary ensembles, the most natural 
examples are the Wigner matrix ensembles \cite{W},
 i.e., random matrices with independent 
identically distributed entries.   The edge universality for these ensembles 
 was proved by  Soshnikov \cite{Sosh}
using  the moment method; the bulk universality remained unknown due
 to a lack of method 
to analyze local spectral properties of large matrices  inside the spectrum.  
For  ensembles  of the form 
\be
\wh H+ a V,
\label{HaV}
\ee
where $\wh H$  is a Wigner matrix,
$V$ is an independent standard GUE matrix
and $a$ is a positive constant of order one (independent of $N$), 
 the bulk  universality was proved by Johansson  \cite{J}.
(Strictly  speaking, the  range of the parameter  $a$ in \cite{J} depends
on the energy $E$.  This restriction was later removed
by  Ben Arous and P\'ech\'e
\cite{BP}, who also extended this approach to Wishart ensembles).

The approach of \cite{J} is partly based on the asymptotic  analysis of an
explicit formula by Br\'ezin-Hikami  \cite{BH}  for 
the correlation functions of the eigenvalues of $\wh H+ a V$. 
This matrix can also be generated by a stochastic flow
\[
s \to  \wh H+ \sqrt{s} V, \quad s>0,
\] 
and  the evolution of the eigenvalues
is given by the  Dyson Brownian motion \cite{Dy}.
The  result of \cite{J, BP}  thus states that
the bulk universality holds for  times of order one.
The  eigenvalue  distribution of GUE is in fact the invariant measure of 
Dyson Brownian motion.
(Rigorously speaking,  the Brownian motion has to be replaced by an 
Ornstein-Uhlenbeck  process,
 but we will neglect this subtlety.)  It is thus  tempting  to derive the
 universality of $\wh H+ \sqrt{s} V$ via  the convergence 
to equilibrium. We have recently carried out this approach \cite{ERSY}
and the key observation  is that the sine kernel, as a 
property of local statistics,  depends almost 
exclusively on the convergence to local equilibrium.
  With this method we have reduced the necessary time to
  $N^{-1+\xi}$, for any $\xi>1/4$ in \cite{ERSY}.
 Note that the relaxation time to local equilibrium is $N^{-1}$; 
the additional exponent $\xi$ is due to technical reasons. 

{F}rom the stochastic calculus, one can see that 
the typical distance between the corresponding eigenvalues of  
$\wh H+ \sqrt s  V$ and $\wh H$ is of order  $(s/N)^{1/2}$.  
Thus the bulk universality 
of $\wh H$ would hold if we 
could  prove the Dyson sine kernel for time 
$s  \ll 1/N$.  On the other hand, for time smaller than $1/N$, 
 the eigenvalues 
do not move in the scale $1/N$ and the dynamical consideration 
seems to be pointless.
In this paper, we provide  an approach to address the comparison
 of eigenvalues between  $\wh H+ \sqrt s  V$ and $\wh H$. 
To describe the idea,  we now introduce the notations.

{F}ix $N\in\NN$ and we consider a Hermitian matrix ensemble 
of $N\times N$ matrices $H=(h_{\ell k})$ 
with the normalization 
\be
   h_{\ell k} = N^{-1/2} z_{\ell k}, \qquad z_{\ell k}=   x_{\ell k}
  +i y_{\ell k},
\label{scaling}
\ee
where $x_{\ell k}, y_{\ell k}$ for $\ell<k$ are independent,
identically distributed random variables
with distribution $\nu$ that has
zero expectation and  variance $\frac{1}{2}$. The diagonal
elements are real, i.e. $y_{\ell\ell}=0$ 
and $x_{\ell \ell}$ are also i.i.d.  with distribution $\wt \nu$
that has zero expectation and  variance one.
The diagonal elements are independent
from the off-diagonal ones.

Suppose the real and imaginary parts of the offdiagonal 
matrix elements evolve according to 
 the  Ornstein-Uhlenbeck (OU) process 
\be\label{dy}
\partial_{t} u_t =  L u_t, \quad 
  L  = \frac{1}{4}\frac{\pt^2}{\pt x^2}
- \frac{ x}{2} \frac{\pt}{\pt x}
\ee
with the reversible  measure $\mu(\rd x) = e^{-x^2} \rd x$
and initial distribution $u_0=u$
(strictly speaking, a differently normalized OU
process is used for the diagonal elements but we omit this detail here).
Under this process, the matrix evolves as
$$
  t\to e^{-t/2}\wh H + (1-e^{-t})^{1/2}V
$$
and the expectation and variance of the matrix entries remain constant.
Notice for time $t$ small, $t \approx a^2$ when compared with \eqref{HaV},
after a trivial rescaling.

The initial distribution of all the matrix elements is
 $F \,\rd\mu^{\otimes n}= ( u\; \rd\mu)^{\otimes n} $
with $n=N^2$. Let $\cL$ be the generator on the product space and
$e^{t\cL} := (e^{tL})^{\otimes n}$  be the dynamics of the OU process
for all the matrix elements.   The
joint probability distribution of the matrix elements  at  time $t$ is then
given by 
\[
F_t \rd \mu^{\otimes n}:=
e^{t\cL} u^{\otimes n} \; \rd\mu^{\otimes n} = (e^{t L} u )^{\otimes n} \; 
\rd \mu^{\otimes n}.
\]
Suppose that for some $t$ small, say, $t = N^{-1+\lambda}$
with $\lambda>0$, we know the local
eigenvalue correlation 
function w.r.t. $F_t$.  Let   
\[
Var(F, F_t) = \int |F- F_t| \rd \mu^{\otimes n}
\]
be the total variation norm between $F_t$ to $F$. 
In order to approximate the correlation functions of $F$ by $F_t$
in a weak sense (tested against bounded observables), 
we need $Var(F, F_t) \to 0$. 
Heuristically, $Var(F, F_t) \sim t N^2$ and this requires that 
$t \ll N^{-2}$ which is far from the time scale  
$t\ge N^{-1+\xi}$ for which the sine kernel has been proven in 
\cite{ERSY}. For observables on short scales, an effective speed of
convergence for the total variation is needed. For example, to
test a local observable with two variables in scale $1/N$, as in the
case of the Dyson sine kernel, one has to prove $Var(F, F_t) = o(N^{-2})$.

Although the heuristic bound  $Var(F, F_t) \sim t N^2$ can be improved to 
$Var(F, F_t) \sim tN$, further improvement seems to be impossible.
Thus we are unable to obtain even the weaker  bound $Var(F, F_t) = o(1)$
 for $t > 1/N$.  
The main observation in the current paper is that, while we
 cannot compare $F$ with 
$F_t$, it suffices to prove the existence of some function $G$ for which 
the  correlation  functions with respect to  $e^{t\cL}G$ can be 
computed for $t\ge N^{-1+\lambda}$
and $Var(F,  e^{t\cL}G) = o(N^{-2})$.  Since the necessary input
  to compute the correlation 
functions is the validity of the semicircle law on short scales, 
 which  we  have proved  for a wide class
of distributions $\nu$ in \cite{ESY1, ESY2, ESY3}, the choice of $G$ is 
essentially dominated by the condition 
$Var(F,  e^{t\cL}G) = o(N^{-2})$. Note that $G$ itself may depend on $t$.
Since $F = e^{t \cL} (e^{-t \cL} F)$, 
we could, in principle, choose  $G = e^{-t \cL} F = [e^{-tL}]^{\otimes n}F$.
 But the diffusive dynamics cannot be reversed 
besides  a very special class of initial data $G$. However, 
 we only have to approximately 
reverse the dynamics 
and the choice   $G_t = \big[ 1-tL+\frac{1}{2}t^2L^2\big]^{\otimes n} F$ 
turns out to be sufficient. In this
case, $e^{t\cL}G_t - F = O(N^2t^3)$ and we will show that 
\be
    \big|Var(e^{t\cL}G_t, F)\big|^2
\le \int \frac{|e^{t\cL}G_t-F|^2}{e^{t\cL}G_t}\rd
 \mu^{\otimes n} = O(t^6N^2).
\label{varg}
\ee
Furthermore,  under some mild regularity condition on $F$,
$G_t$ is in the class for which we 
can establish the local semicircle law \cite{ESY3}. We will
call this argument the {\it method of time reversal.}

\medskip

We now summarize the assumptions on the initial distribution. 
Let the probability measure of the real and imaginary
parts of the off-diagonal matrix elements be of the
form  
$$
   \nu(\rd x) = e^{-U(x)}\rd x=u(x) \mu(\rd x) = e^{- V(x)} e^{- x^2}\rd x 
$$
with the real function $V(x)= U(x) -x^2$ 
and similarly for the diagonal
elements $\wt\nu(\rd x) = e^{-\wt U(x)}\rd x$,  $\wt V(x)= 
\wt U(x) -\frac{1}{2}x^2$.
Suppose that $V \in C^6 (\bR)$ and the derivatives satisfy
\be
\sum_{j=1}^6 |V^{(j)}(x)| \le C (1+x^2)^k
\label{cond1}
\ee
for some $k\in \NN$ 
and 
\be
\nu(x) \le C' e^{- \delta |x|^2}
\label{cond2}
\ee
with some constants $\delta>0$, $C$ and $C'$. 
In Section \ref{sec:relax} we explain how to relax
this latter  condition to exponential decay,
\be
  \nu(x) \le C' e^{-C|x|}
\label{cond2relax}
\ee
with some constants $C, C'$ (in fact, some
high power law decay is sufficient). We 
assume that the first moment of $\nu$ is zero
and the variance is $\frac{1}{2}$
\be
\int x \, \rd \nu(x)=0 ,\qquad \int x^2 \rd \nu(x)=\frac{1}{2}.
\label{cond3}
\ee
 We assume the  conditions \eqref{cond1}, \eqref{cond2}
and \eqref{cond3} for $\wt V$ as well with the
variance changed to 1.

Let $p_N(x_1, x_2, \ldots , x_N)$ denote
the probability  density of eigenvalues and for any
$k=1,2,\ldots, N$,
let 
\be
  p^{(k)}_N(x_1, x_2,\ldots x_k):= 
 \int_{\bR^{N-k}} p_N(x_1, x_2, \ldots , x_N)\rd x_{k+1}\ldots \rd
x_N
\label{corrfn}
\ee
be the $k$-point correlation function.
With our choice of the variance of $\nu$, the density $p^{(1)}_N(x)$
is supported in $ [- 2,   2]+o(1)$ and in the $N\to\infty$ limit
it converges to the Wigner semicircle law given by the density
\be
 \varrho_{sc}(x)= \frac{1}{2\pi} \sqrt{(4-x^2)_+}\; .
\label{def:sc}
\ee

\begin{theorem}\label{mainthm}
Let the probability measure of the matrix elements
satisfy conditions \eqref{cond1}, \eqref{cond2} and \eqref{cond3}.
 Then for any $u$ with $|u|< 2$ and 
for any compactly supported and bounded observable
$O\in L^\infty_c(\bR^2)$ we have
\be
\begin{split}
\label{maineq}
\lim_{N\to \infty} \int_{\bR^2} O(\al,\beta) \frac{1}{[\varrho_{sc}(u)]^2}
p^{(2)}_N\Big(u + 
\frac{\al}{N\varrho_{sc}(u)}, &
  u + \frac{\beta}{N\varrho_{sc}(u)}\Big)
  \rd \al \rd \beta \\
& = \int_{\bR^2} O(\al,\beta) \Big[1-
\Big(\frac{\sin \pi(\al-\beta)}{\pi(\al-\beta)}\Big)^2\Big] \rd \al \rd \beta.
\end{split}
\ee
\end{theorem}

\begin{remark}{\rm 
With similar methods we can also
prove that the higher order rescaled correlation functions,
$$
    \frac{1}{[\rho_{sc}(u)]^k}
\;p^{(k)}_N\Big(u+ \frac{ a_1 } { \rho_{sc}(u) N},
u+\frac{ a_2}
{ \rho_{sc}(u) N }, \ldots, u+ \frac{ a_k } { \rho_{sc}(u) N}
 \Big),
$$
converge in the weak sense to $\mbox{det}\big( f(a_i-a_j)\big)_{1\le
i, j\le k}$ where $f(\tau)= \frac{\sin \pi \tau}{\pi \tau}$,
however this statement requires more regularity conditions on $V$.
The proof of the sine kernel for $e^{tL}G_t$ immediately implies
the convergence of the higher order correlation functions
with respect to the evolved measure. To conclude for
the higher order correlation functions with respect to $F$, however, one
needs to improve the accuracy in \eqref{varg}. This can
be achieved by approximating
the backward evolution $e^{-t\cL}$ to a higher order. For example,
using
$G_t=\big[ 1- tL + \frac{1}{2!} (-tL)^2 -\ldots \frac{1}{(m-1)!}
(-tL)^{m-1}\big]^{\otimes n}F$, will improve the bound \eqref{varg}
to $t^{2m}N^2$, modulo $N^\e$ corrections, if $V$ is
$2m$-times differentiable with bounds similar to \eqref{cond1}.
}
\end{remark}

\begin{remark}{\rm  With the same method, the condition 
that $V \in C^6 (\bR)$ in Theorem \ref{mainthm} can be relaxed
 to $V \in C^{4+\eps} (\bR)$ for any $\e>0$.
Heuristically, this can be seen by observing that, with 
$F= v^{\otimes n}$ and $G_t = v_t^{\otimes n} = [1-tL +\frac{1}{2} t^2 L^2]^{\otimes n} F$, one can estimate the difference $|e^{tL} v_t - v| \leq O(t^{2+\eps}
 L^{2+\eps} v)$, which, compared with the estimate $O(t^3 L^3 v)$ 
used in \eqref{varg} gives less decay (still enough to deliver the 
result of Theorem \ref{mainthm}), but requires less regularity of $v$ 
(only $4+2\eps$ derivatives). A rigorous proof of this fact can be 
obtained by using part of the evolution $e^{t\cL}$ to regularize, 
on the scale $t$, the initial density.}
\end{remark}

We now state our result concerning  the eigenvalue gap distribution. 
For any $s>0$ and $|u|<2$ we 
define the density  of eigenvalue pairs  with distance less than
 $s/N\varrho_{sc}(u)$ in the vicinity of $u$ by
\be
\Lambda (u; s, x) = 
\frac {1}{2  N t_N \varrho_{sc}(u)}
 \# \Big\{ 1\le j \le N-1\,: \; x_{j+1} - x_j \le
 \frac{s}{N\varrho_{sc}(u)}, \; |x_j-u| \le t_N  \Big\}
\label{def:Lambda}
\ee
where $t_N = N^{-1+ \delta}$ for some $0< \delta< 1$.

\begin{theorem}\label{mainthm2}
Suppose  the probability measure of the matrix elements
satisfies  conditions \eqref{cond1}, \eqref{cond2} and \eqref{cond3}.
Let $\cK_\alpha$ be the operator acting 
on $L^2((0, \alpha))$ with kernel $\frac {\sin \pi(x-y)}
{\pi(x-y)}$.
 Then for any $u$ with $|u|< 2$ and for any $s> 0$ we have
 \be\label{maineq2}
\lim_{N\to \infty}  \E \, \Lambda (u; s, x)
= \int_0^s  p(\alpha)\; \rd \alpha  , \qquad p(\alpha) = 
\frac {\rd^2} {\rd \alpha^2}  \det (1 - \cK_\alpha),
\ee
where $\det$ denotes the Fredholm determinant of the
compact operator $1-\cK_\al$.
\end{theorem}
As a corollary of Theorem \ref{mainthm2}, one can easily show that the
probability to find no eigenvalue in the
 interval $[u,u+\al/(\varrho_{sc} (u_0) N)]$, 
after averaging in an interval of size $N^{-1+\delta}$ around 
$u_0 \in (-2,2)$, is given by
 $\det (1 - \cK_\alpha)$, same as in the case of GUE
 (see, e.g., \cite{D}).  Note that assuming more regularity 
on the exponent of the density $u(x) = e^{-U(x)}$, we can get 
a better bound on the convergence rate 
(by approximating the backwards evolution $e^{-t\cL}$ to a higher order) 
and avoid therefore the averaging over $u$.

\medskip

 The proof of Theorem \ref{mainthm} and  
\ref{mainthm2} consists of two main parts. In Section \ref{sec:rev} we 
 prove the approximation \eqref{varg} under precise conditions
on the initial distribution $u=e^{-V}$.
In Section \ref{sec:timeevolved} we  prove 
the sine kernel
for the distribution $e^{t\cL}G_t$ with $t=N^{-1+\lambda}$ for any $\lambda>0$, 
which is the optimal  time scale for such a result.   Our approach is to recast the formula 
for the correlation function in \cite{J}, which  becomes  unstable  for  $ t \ll 1$, into a more symmetric form (Proposition \ref{prop:rep})
so that it is stable 
for all time up to $t=N^{-1+\lambda}$.  The saddle point analysis can then be achieved with the local semicircle law from \cite{ESY3}. 
Finally, we complete the proofs
of the main theorems in Section \ref{sec:mainthm}.

\medskip

The method of time reversal described 
previously is  very general and should be applicable to a wide
range of models. More significantly, it explains the {\it origin} 
of the universality, i.e., the
universality
comes from the ``time reversal''. To summarize, the universality
 consists of the following
observations:
(1) The local statistics  are determined by the local equilibrium 
measures. (2) The relaxation to
local
equilibria takes place in a short time. (3) The original 
distribution can be well-approximated by the distribution
of the Dyson Brownian motion for a short time with 
initial data given by an approximate
inverse flow. To implement this scheme,
a key input is to estimate the fluctuations
of the empirical density of eigenvalues in short scales.

\medskip

Shortly after this manuscript appeared on the arXiv, 
we learned that our main result was also obtained by Tao and Vu in
\cite{TV} under essentially no regularity conditions on the initial 
distribution $\nu$ provided the third moment of $\nu$ vanishes. 
Some partial results for the Gaussian orthogonal ensembles are also
obtained and we refer the reader to the preprint for more details.

\medskip

{\it Conventions.} 
We will use the letters $C$ and $c$ to denote general constants whose
precise values are irrelevant and they may change from line to line.
These constants may depend on the constants in \eqref{cond1}--\eqref{cond3}.

\section{Method of Time Reversal}\label{sec:rev}

Recall the  Ornstein-Uhlenbeck process from \eqref{dy} 
with the reversible  measure $\mu(\rd x) = \mu(x)\rd x=e^{-x^2} \rd x$. 
Let $u$ be a positive density with respect to $\mu$, 
i.e. $\int u \rd \mu=1$
and we write $u(x)=\exp (-V(x))$.

\begin{proposition}\label{meascomp}
Let $V$ satisfy the conditions \eqref{cond1}, \eqref{cond2} 
with some $k$ and \eqref{cond3}.
 Let $\lambda>0$ be sufficiently small and   $t= N^{-1+\lambda}$.
Define a cutoff initial density as
\[
 v (x):= e^{-  V_c(x) } ,\qquad
  V_c (x):= V(x) \theta((x-c_N)N^{-\lambda/4k})+d_N,
\]
where $\theta$ is a smooth cutoff function satisfying 
$\theta(x) = 1$ for $|x|\le 1$ and $\theta(x) = 0$ for $|x| \ge 2$
and $c_N$ and $d_N$ are chosen such that $v(x)\rd\mu(x)$ is a probability
density with zero expectation.
Denote
  $\cL=L^{\otimes n}$, $ F=  u^{\otimes n}$ and $ F_c= v^{\otimes n}$
with $n=N^2$.

(i) We have
\be\label{FFtilde}
  \int  \left |  { F_c }  - F \right |  
\rd \mu^{\otimes n}   \le C\; e^{- cN^{c}}.
\ee
with some $c>0$ depending on $k$ and $\lambda$.

(ii)  $g_t:= (1-tL+\frac{1}{2}t^2L^2)v$ 
is a probability measure with respect to $\rd\mu$
and for $G_t:=[g_t]^{\otimes n}$ we have
\be
    \int  \frac{\big|e^{t\cL}G_t- F_c\big|^2}{e^{t\cL}G_t} 
\rd \mu^{\otimes n} \le CN^2t^{6-\lambda} \le CN^{-4+8\lambda},
\label{FF}
\ee
where $C$ depends on $\lambda$
 and on the constants in \eqref{cond1}, \eqref{cond2}.
\end{proposition}

In the formulation of this
 proposition we have not taken into account that in our application the
diagonal elements of the matrix  evolve under a differently normalized
OU process with generator $\wt L = \frac{1}{2}\pt_x^2 - \frac{x}{2}\pt_x$
with invariant measure $e^{-x^2/2}\rd x$.  This modification 
is only notational and does not affect the validity of the
estimates \eqref{FFtilde} and  \eqref{FF}.

\medskip
{\it Proof.} {F}rom condition \eqref{cond2} 
the estimate \eqref{FFtilde} follows directly by noting
that the constants $c_N$ and $d_N$
are subexponentially small in $N$.
For the proof of \eqref{FF}, we 
 first control the evolution of each matrix element
under the OU process \eqref{dy}. 
 We assume that for the initial density $v$ 
\be
Lv (x)  \le  A_1  v (x) , 
\qquad   L^2 v(x)\ge - A_2  v(x), \qquad
|L^3v(x)|\le A_3 v(x)
\label{AB}
\ee
hold with some constants positive $A_1,  A_2$ and $A_3$.
 Set $g_t = (1-tL+\frac{1}{2}t^2L^2)v$
 for some $t>0$ and note
that $g_t$ is  a probability density with respect to $\mu$
if 
\be 
tA_1 + \frac{t^2}{2}A_2\le 1.
\label{tAA}
\ee
Define
$$
  v_t = e^{tL}g_t= e^{tL} \Big(1-tL+\frac{1}{2}t^2L^2\Big)v,
$$
then
$$
  \pt_t v_t = \frac{1}{2} t^2 L^3 e^{tL}v.
$$
Note that by the monotonicity preserving property of the
Ornstein-Uhlenbeck kernel and by \eqref{AB}, we have
\be
   e^{sL} L^3 v \le A_3 e^{sL} v \le A_3 e^{sA_1} v, \qquad s\ge 0.
\label{derest}
\ee
Here we used the fact that $e^{sL}v\le e^{sA_1}v$ under the
first condition in \eqref{AB}, which follows from integrating
the inequality
$$
  \frac{\rd}{\rd s} e^{sL}v = e^{sL}Lv \le A_1 e^{sL}v.
$$ 

In particular
\be
   v_t = v + \frac{1}{2}\int_0^t s^2 L^3 e^{sL}v \; \rd s
 \ge v\Big(1- \frac{1}{6}t^3 A_3 e^{tA_1}\Big) \ge 
\frac{1}{2}v,
\label{uut}
\ee
assuming \eqref{tAA} and
\be
    t^3 A_3 \le 1.
\label{tAA1}
\ee
Then
\be
\begin{split}\label{uu1}
   \int \frac{(v-v_t)^2}{v_t} \;
 \rd\mu & = \int v_t^{-1} \Big[ \int_0^t \rd s \; \frac{1}{2}s^2
   L^3 e^{sL} v\Big]^2 \rd\mu \\
 & \le \frac{t^5}{20}
   \int_0^t \int v_t^{-1} \big[  e^{sL}L^3 v\big]^2\rd\mu\rd s \\
 & \le \frac{t^5}{10}\int_0^t\int v^{-1}  \big[ L^3 e^{sL} v\big]^2\rd\mu\rd s
 \\
 & \le \frac{1}{10}\, A_3^2 t^6 e^{2tA_1}\le e^{CA_3^2 t^6}-1, 
\end{split}
\ee
where we used \eqref{uut}, \eqref{derest} and finally \eqref{tAA}.

\medskip

Now we consider the evolution of the product density
$F_c= v^{\otimes n}$, note that $\int F_c \; \rd\mu^{\otimes n}=1$.
 Applying the same procedure to each variable,
 we have 
\be\label{CA}
  \int  \frac { (e^{t\cL}G_t-F_c)^2 } {e^{t\cL}G_t } \; \rd \mu^{\otimes n}  
\le  e^{CA_3^2 t^6n}-1\le CA_3^2 t^6n \, 
\ee
as long as $A_3^2 t^6n$ is bounded. 
In our application  $n=N^2$, thus \eqref{CA} will imply \eqref{FF} 
 provided that 
\be
  A_3 \le Ct^{-\lambda/2}
\label{A3}
\ee
which will also guarantee \eqref{tAA1}. It is straightforward
to check that the density $v(x)$ satisfies \eqref{AB} with 
constants $A_j$ subject to \eqref{tAA} and \eqref{A3}.
This completes the proof. \qed

\section{Sine kernel for the time evolved measure}
\label{sec:timeevolved}

We use the contour integral representation for
the correlation functions of the eigenvalues of
a matrix of the form $H=\wh H + aV$, where $V$ is a GUE
matrix \cite{BH, J}. We will apply this result for
the matrix
\be
   e^{t\cL}G_t = e^{-t/2}\big[ G_t + (e^t-1)^{1/2} V\big]
\label{resc}
\ee
where, apart from a trivial
prefactor $e^{-t/2}$, 
 $G_t$ plays the role of $\wh H$ and $a= (e^t-1)^{1/2}\approx t^{1/2}$.
 In order to be able to use
the formula given in Proposition 1.1 of \cite{J}
to analyze  $H=\wh H + aV$,
we rescale the variance of $\rd\nu$ from $\frac{1}{2}$ to 
$\frac{1}{8} +\frac{1}{2}a^2$
 which changes the
 semicircle law for $H=\wh H +aV$  to
\be
  \varrho(u) : = \frac{2}{\pi(1+4a^2)}\sqrt{(1+4a^2-u^2)_+}.
\label{def:varrho}
\ee
In  particular, the support   changes
from $[-2,2]$ to $[-\sqrt{1+4a^2},\sqrt{1+4a^2}]$.
Since
eventually $a$ will go to zero, the condition $|u|< 2$
in Theorem \eqref{mainthm}
to be away from the spectral edge changes to the
condition $|u|<1$ which we assume in the sequel. 
The  semicircle law for $\wh H$ will also change
from the one given
in  \eqref{def:sc} to
\be
  \varrho_{sc}(v) := \frac{2}{\pi}\sqrt{(1-v^2)_+}.
\label{def:sc1}
\ee
In the rest of this Section we will use \eqref{def:sc1}.
The main result of this section is

\begin{proposition}\label{sinjoh}
Let $\wt p^{(m)}_{N}$ be the $m$-point eigenvalue
correlation function for the ensemble $\wh H + aV$ defined above 
and let $O: \bR^m\to \bR$ 
be a compactly supported bounded observable function. Then
for any $|u|<1$ and $a:=  N^{-1/2+\lambda/2}$ we have
\be
\begin{split}
\lim_{N\to \infty} \int_{\bR^m} O(\al_1, \ldots, \al_m) & 
 \frac{1}{[\varrho(u)]^m}
 \wt p^{(m)}_{N}\Big(u +  \frac{\al_1}{N\varrho(u)},\ldots,
  u + \frac{\al_m}{N\varrho(u)}\Big)
   \rd \al_1\ldots \rd \al_m \\
& = \int_{\bR^m} O(\al_1, \ldots, \al_m)
\det\Big(\frac{\sin \pi(\al_i-\al_j)}{\pi(\al_i-\al_j)}\Big)_{i,j=1}^m
 \rd \al_1\ldots \rd \al_m.
\end{split}
\ee
\end{proposition}

{\it Proof.}
Using Proposition 1.1 of \cite{J}, the (symmetrized) distribution of the
eigenvalues $x=(x_1, \ldots, x_N)$ of $H=\wh H + aV$ for any fixed $\wh H$
is given by
\be
    q_S(x,y) : = \frac{1}{(2\pi S)^{N/2}} \frac{\Delta_N(x)}{\Delta_N(y)}
  \mbox{det}\big( e^{-(x_j-y_k)^2/2S}\big)_{j,k=1}^N,
\label{def:qs}
\ee
where $y=(y_1, \ldots y_N)$ is the eigenvalues of the
Wigner matrix $\wh H$ with the choice of $S= a^2/N$.
Note that 
\be\label{allsum}
\begin{split}\int_{\bR^m}& O(\al_1, \ldots, \al_m)  
  \frac{1}{[\varrho(u)]^m}
 \wt p^{(m)}_{N}\Big(u +  \frac{\al_1}{N\varrho(u)},\ldots,
  u + \frac{\al_m}{N\varrho(u)}\Big)
   \rd \al_1\ldots \rd \al_m \\
& =  \wh\EE \int_{\bR^N}
  \sum_{i_1, i_2, \ldots, i_m=1}^N O\Big( N\varrho(u)(x_{i_1}-u),
\ldots,     N\varrho(u)(x_{i_m}-u)\Big) q_S(x,y)\rd x_1 \ldots \rd x_N,
\end{split}
\ee
where $\wh \EE$ denotes the expectation is w.r.t. the $\wh H$ ensemble.
Since $O$ is bounded and the sum contains $N^m$ terms,
we thus need to compute the limit of  the
correlation functions of $q_S(x,y)$
in the $x=(x_1, \ldots, x_N)$ variables for a large set $\cY_N\subset \bR^N$
of fixed $y=(y_1, \ldots , y_N)$ so that 
$$
   \wh \PP (y(\wh H)\not\in\cY_N)  = o(N^{-m}).
$$
where  $y(\wh H)
=(y_1(\wh H), \ldots, y_N(\wh H))$ are the eigenvalues
of the Wigner
matrix $\wh H$. We will choose $\cY_N$ to be the event that the points
$y=(y_1, \ldots , y_N)$ follow the semicircle law \eqref{def:sc1}.
The limit of the  
correlation functions of $q_S(x,y)$ will be computed starting from
the next section in Proposition \ref{prop:local}.

More precisely,
let 
\be
\eta := \eta_0t\sqrt{1-u^2}
\label{eta}
\ee
with some sufficiently small $\eta_0<1$ and we set
\be
   \cY_N: = \Big\{ y\in \bR^N\; : \; 
 \; \sup_{Im z\ge\eta}
 \Big|  \frac{1}{N}\sum_j \frac{1}{z-y_j} - \int 
  \frac{\varrho_{sc}(r)\rd r}{z-r}\Big|\le N^{-\lambda/4} 
  \;\; \mbox{and} \quad\sup_j |y_j|\le K\Big\}
\label{defY}
\ee
for some large constant $K$. 

By Theorem 3.1 of \cite{ESY3}
we then have
\be
 \wh\PP (y(\wh H)\not\in\cY_N) \le Ce^{-cN^{\lambda/4}}
\label{good}
\ee
(after taking the
supremum over all energies, which can be controlled
taking energies on a grid of spacing $\eta$).
Note that the variance of the matrix elements
in \cite{ESY3} was different (see remark at the beginning
of Section \ref{sec:con}) but this does not change the
estimates. The condition {\bf C1)} of \cite{ESY3}
on the Gaussian decay for
the initial density $g_t\mu =(1- tL+\frac{1}{2}t^2L^2)v\mu$
is clearly satisfied by \eqref{AB} and \eqref{cond2}.
Combining the estimate \eqref{good} with Proposition \ref{prop:local}
and with the argument after \eqref{allsum}, we have proved 
Proposition \ref{sinjoh}.\qed

\medskip

\subsection{Contour integral representation of the correlation function}
\label{sec:con}

We compute the correlation functions of $q_S(x;y)$ 
in $x$, for any fixed $y\in\cY_N$:
\be
  \wt p_{N,y, S}^{(m)}(x_1, \ldots , x_m) 
= \int_{\bR^{N-m}} q_S(x_1, \ldots ,x_N; y)
   \rd x_{m+1} \ldots \rd x_N.
\label{tildecorr}
\ee
Note that this definition of the correlation 
functions  differs from
the definition of $R_m^N$ given in \cite{J}; the relation being
$$
    R_m^N(x_1, \ldots, x_m; y) = \frac{N!}{(N-m)!}\wt p_{N,y, S}^{(m)}(x_1,
 \ldots , x_m) .
$$
The following representation is based on the formula in \cite{J}, but it is more stable and suitable for analysis 
for very short time. 

\begin{proposition}\label{prop:rep} The correlation functions can be represented as
\be R_m^N (x_1, \dots ,x_m ; y) = \mbox{det}
\big( \cK_N^S(x_i, x_j; y)\big)_{i,j=1}^m,
\label{def:R}
\ee
where
\be
\begin{split}\label{ck}
  \cK_N^S (u,v;y)= & \frac{1}{(2\pi i)^2 (v-u)S}
  \int_\gamma \rd z\int_\Gamma \rd w (e^{-(v-u)(w-r)/S} -1) \prod_{j=1}^N 
  \frac{w-y_j}{z-y_j} \\
 &  \times \frac{1}{w-r}\Big( w-r+z-u - S\sum_j \frac{y_j-r}{(w-y_j)(z-y_j)}\Big)
  e^{(w^2-2uw -z^2+2uz)/2S},
\end{split}
\ee
where $r\in \bR$ is arbitrary and $\gamma= \gamma_+\cup\gamma_-$ is 
the union of two lines $\gamma_+:s\to -s + i\om$ and
 $\gamma_-:s\to s-i\om$ ($s\in \bR$)
for any fixed $\om>0$ and $\Gamma$ is $s\to is$, $s\in \bR$.
\end{proposition}

We note that $\Gamma$ 
can be shifted to any vertical line since the
integrand is an entire function in $w$ and has a Gaussian decay
as $|Im \; w| \to \infty$. 
 The constants $r \in \bR$ and $\om>0$  (appearing in the
definition of the contour $\gamma$ in $K_N$)  can be arbitrary
and will be specified later.

\medskip

{\it Proof of Proposition \ref{prop:rep}.}
{F}rom Eq. (2.18) in \cite{J}, we have 
\be R_m^N (x_1, \dots ,x_m ; y) = \mbox{det}
\big( K_N^S(x_i, x_j; y)\big)_{i,j=1}^m,
\label{R2}
\ee
with 
\[ K_N^S (u,v;y) = K_N^S(u,v):= \frac{e^{(v^2-u^2)/2S}}{(2\pi i)^2 S} \int_{\wt\gamma} \rd z 
\int_{\Gamma_L} dw \, e^{(w^2-2wv -z^2+2zu)/2S} \frac{1}{w-z} 
\prod_{j=1}^N \frac{w-y_j}{z-y_j}\, , \]
where $\wt\gamma$ is a contour around all the $y_j$, $j=1, \dots ,N$, 
and $\Gamma_L$ is the vertical line $\R \ni s \to L +is$, for a fixed 
$L$ so large that $\wt\gamma$ and $\Gamma_L$ do not intersect. Eq. (\ref{R2}) 
remains invariant if we replace $K_N$ by 
\[ 
\cK^S_N (u,v) = e^{r(v-u)/S} e^{(u^2-v^2)/2S} K^S_N (u,v) = \frac{1}{(2\pi i)^2 S}
 \int_{\wt\gamma} \rd z \int_{\Gamma_L} \rd w \,  
\frac{e^{r(v-u)/S}}{w-z} \, e^{(H_v(w) - H_u (z))/S} 
\] 
for arbitrary $r\in \bR$. Here we defined 
$$
   H_v (w) := \frac{w^2}{2} -vw +S \sum_{j=1}^N \log (w-y_j)\, .
$$ 
The change of variables $w= (1-\beta) r + \beta w'$, $z=(1-\beta) r +\beta z'$ 
leads to 
\[ \cK^S_N (u,v) := \frac{\beta}{(2\pi i)^2 S} \int_{\wt\gamma} \rd z'
 \int_{\Gamma_L} \rd w' \, \frac{e^{r(v-u)/S}}{w-z} \, 
e^{(H_v((1-\beta)r+\beta w') - H_u ((1-\beta) r + \beta z'))/S} 
\]
for every $\beta$. Taking the derivative in $\beta$ at $\beta=1$,
and removing the primes from the new integration variables, we find the identity
\[ 0 = \cK^S_N (u,v) + \frac{1}{(2\pi i)^2 S} \int_{\wt\gamma} \rd z \int_{\Gamma_L} \rd w 
\,  e^{r(v-u)/S} \, e^{(H_v(w) - H_u (z))/S} \, \frac{1}{S} 
\left[ \frac{(w-r) H'_v (w) - (z-r) H'_u (z)}{w-z} \right] .
\]
Using that $H'_v (w) = w-v + S \sum_{j=1}^N 1/(w-y_j)$, we find 
\[  \frac{(w-r) H'_v (w) - (z-r) H'_u (z)}{w-z} = 
\frac{(w-r)(u-v)}{w-z} + (w-r) \frac{H'_u (w) - H'_u (z)}{w-z} +H'_u(z) 
\]
and thus
\[ 
\begin{split} 
0 = \, &\cK^S_N (u,v) + \frac{(u-v)}{(2\pi i)^2 S} \int_{\wt\gamma} \rd z 
\int_{\Gamma_L} \rd w \,  e^{r(v-u)/S} \, \frac{w-r}{S(w-z)} 
e^{(H_v(w) - H_u (z))/S} \, 
\\ &+ \frac{1}{(2\pi i)^2 S} \int_{\wt\gamma} \rd z \int_{\Gamma_L} \rd w \, 
 e^{r(v-u)/S} \, e^{(H_v(w) - H_u (z))/S} \, \frac{1}{S} 
\left[ (w-r) \frac{H'_u (w) - H'_u (z)}{w-z} + H'_u (z) \right]\,. 
\end{split}
 \]
The second term on the r.h.s. is just $(v-u) \frac{\partial}{\partial v}
 \cK_N (u,v)$. Therefore
\[ 
\begin{split} 
 \frac{\partial}{\partial v} & \left[ (v-u) \cK^S_N (u,v)\right] \\ 
= \; &  \frac{-1}{(2\pi i)^2 S} \int_{\wt\gamma} \rd z \int_{\Gamma_L} \rd w \, 
e^{r(v-u)/S} \, e^{(H_v(w) - H_u (z))/S} \, \frac{1}{S} 
\left[ (w-r) \frac{H'_u (w) - H'_u (z)}{w-z} + H'_u (z) \right] \\ 
= \; &\frac{1}{(2\pi i)^2 S} \int_{\wt\gamma} \rd z \int_{\Gamma_L} \rd w \, 
e^{r(v-u)/S} \, e^{(H_v(w) - H_u ( z))/S} \, \frac{1}{S}
 \left[ w-r +z-u -S \sum_{j=1}^N \frac{y_j -r}{(z-y_j)(w-y_j)} \right] . 
\end{split} 
\]
Integrating back over $v$, starting from $u$, we find that 
\[ 
\begin{split}
 (v-u) \cK^S_N (u,v) = \frac{1}{(2\pi i)^2 S}  \int_{\wt\gamma} \rd z \int_{\Gamma_L} \rd w 
\,& \left(e^{-(w-r)(v-u)/S} - 1 \right) \, e^{(w^2 - 2uw - z^2 + 2uz)/2S} 
\prod_{j=1}^N \frac{w-y_j}{z-y_j} 
\\  &\times  \frac{1}{(w-r)} \left[ w-r +z-u -S \sum_{j=1}^N 
\frac{y_j -r}{(z-y_j)(w-y_j)} \right] \,. 
\end{split} \]
At this point the contours of integration can be modified; since 
the singularity $1/(w-z)$ has been removed, they are now allowed to cross.
 This completes the proof of the  proposition. \qed

\begin{proposition}\label{prop:local} Let $\kappa>0$. 
For any sequence $y=y^{(N)}\in \cY_N$ with the choice 
$S= N^{-2+\lambda}$ we have
\be
   \lim_{N\to \infty} \frac{1}{N\varrho(u)}
\cK_N^S\Big( u + \frac{\al}{N\varrho(u)}
   , u+ \frac{\beta}{N\varrho(u)} ; y\Big)
   = \frac{\sin \pi(\al-\beta)}{\pi(\al-\beta)}
\label{convpoint}
\ee
uniformly for $|u|\le 1-\kappa$ and for $\al,\beta$ in a compact set.
Moreover, the correlation functions satisfy
\be
    \lim_{N\to\infty} \frac{1}{[\varrho(u)]^{m}} \wt p_{N,y, S}^{(m)}
  \Big( u+\frac{\al_1}{N\varrho(u)}, \ldots , u+\frac{\al_m}{N\varrho(u)}\Big)
= 
  \det \Big( \frac{\sin \pi(\al_i-\al_j)}{\pi(\al_i-\al_j)}
\Big)_{i,j=1}^m,
\label{pmconv}
\ee
uniformly for $|u|\le 1-\kappa$ and for $\al_1, \ldots, \al_m$
 in a compact set.
\end{proposition}

{\it Proof.}  The statement in \eqref{pmconv} follows
directly from \eqref{convpoint} and \eqref{def:R}, so
it is sufficient to prove \eqref{convpoint}.
We will prove \eqref{convpoint} in the form
$$  
 \frac{1}{N\varrho(u)}
\cK_N^S\Big( u^{(N)} 
   , u^{(N)}+ \frac{\tau}{N\varrho(u)} ; y\Big) \to 
 \frac{\sin \pi\tau}{\pi\tau}
$$
for any sequence $u^{(N)}$ with $|u^{(N)} - u_*| \leq C/N$ and for every fixed $u_*$ with $|u_*| <1-\kappa$. In order to get (\ref{convpoint}), we take $u^{(N)} = u_* +\alpha/N\varrho (u_*)$ with $u_* = u$.   

Set 
\be
 \varrho = \varrho(u_*), 
 \quad t=a^2  = N^{-1+\lambda}.
\label{not}
\ee
{F}rom (\ref{ck}), we find 
\be
  \frac{1}{N\varrho}\cK_N\Big(u^{(N)},u^{(N)}+\frac{\tau}{N\varrho}; y\Big)
  = N \int_\gamma \frac{\rd z}{2\pi i}\int_\Gamma \frac{\rd w}{2\pi i}
  h_N(w) g_N(z,w) e^{N(f_N(w)-f_N(z))}
\label{repr}
\ee
with
\be
   f_N(z) = \frac{1}{2t}(z^2-2u^{(N)} z) +\frac{1}{N}\sum_j\log(z-y_j)
\label{def:fN}
\ee
\be
  g_N(z,w) = \frac{1}{t(w-r)}[w-r+z-u^{(N)}] -
\frac{1}{N(w-r)}\sum_j \frac{y_j-r}{(w-y_j)(z-y_j)}
\label{def:gN}
\ee
\be
\begin{split}
   h_N( w)  & =
\frac{1}{\tau} \Big( e^{-\tau (w-r)/t\varrho}
  - 1 \Big)
\label{def:hN}
\end{split}
\ee
with \eqref{not}.
We will need the identity
\be
  g_N(z,w) = \frac{1}{w-r} f_N'(z) + \frac{f'_N(z)-f_N'(w)}{z-w}.
\label{id}
\ee

\subsection{Saddle points}

For brevity, we will drop the superscript and denote $u^{(N)}$ by $u$ in the sequel and we fix $|u|<1$. We first determine the critical points of $f_N$, i.e. we solve
\be
  f_N'(z) = t^{-1}(z-u) +\frac{1}{N}\sum_j\frac{1}{z-y_j} =0.
\label{fNroot}
\ee
This is equivalent to finding the zeros of a polynomial of degree $N+1$.
There are $N-1$ real roots and two complex roots, called $q_N^\pm$, 
that are complex conjugates of each other
$$
   f_N'(q_N^\pm)=0.
$$
 We will work with
$q_N:= q_N^+$, the analysis of the other saddle is
analogous. Clearly $|Re \; q_N |\le K$ for some large $K$.

We can define
$$
   f(z) =\frac{1}{2t}(z^2-2uz) + \int_\bR\varrho_{sc}(y) \log (z-y) \rd y
$$
and instead of \eqref{fNroot}, we can solve
\be
  f'(z)  =  t^{-1}(z-u) + 2(z-\sqrt{z^2-1}) =0.
\label{fprime}
\ee
The solutions of this latter equation (for small $t$) are given by
\be
   q^\pm =\frac{(2t+1)u\pm 2ti\sqrt{1+4t-u^2}}{1+4t} =
   u(1-2t)\pm 2ti\sqrt{1-u^2} +O(t^2),
\label{qsol}
\ee
and thus in particular
$$
   Im (q^\pm) = \pm O(t).
$$
We have
$$
  f''(q) = \frac{1}{t}+2 -\frac{2q}{\sqrt{q^2-1}}
$$ 
and 
$$
 f''(q^\pm) 
  = \frac{1}{t}+2 \pm \frac{2ui}{\sqrt{1-u^2}} + O(t)
$$
where we also used the equation \eqref{fprime} for $q^\pm$.
We set $q=q^+$.

We need to know that $f_N''\ne 0$ at the $q_N$ saddle.
$$
  f_N''(q_N) = \frac{1}{t} - \frac{1}{N}\sum_j \frac{1}{(q_N-y_j)^2}.
$$

It follows from \eqref{defY} that for $y\in \cY$ we have
\be
    \sup_{Im z \ge \eta}
   |f_N^{(\ell)}(z) - f^{(\ell)}(z)|\le \frac{C}{t^{\ell-1}N^{\lambda/4}}
\label{ellgood}
\ee
by contour integration.

\bigskip

We compare $q$ and $q_N$. We have from \eqref{fNroot}
\be
    q_N= F_N(q_N):= u- \frac{t}{N}\sum_j\frac{1}{q_N-y_j} , \qquad Im \; q_N>0
\label{qNeq}
\ee
and
\be  
   q =  F(q):=u -t \int \frac{\varrho_{sc}(y)\rd y}{q-y} =
u - 2t(q-\sqrt{q^2-1}) 
\label{qeq}
\ee
First we show that for the only solution to 
\eqref{qNeq} with positive imaginary part we have
$Im \; q_N\ge \eta$.
 This is a fixed point argument.

Define the compact set
$$
  \Xi :=\Big\{ z\; : \; |Re \; z - u|\le Ct, \; 
  \eta\le Im \; z \le Ct \Big\}
$$
for some large constant $C$.
Since $y\in\cY$, we know that
$$
   \sup_{Im\; z\ge \eta} |F_N(z) - F(z)| \le \frac{Ct}{N^{\lambda/4}}.
$$
For $z\in \Xi$ clearly 
$$
     Re \; F(z)= u + O(t)
$$
and
$$
   Im \; F(z) = 2t\sqrt{1-u^2} + O(t^2)
$$
thus
$$
    Re \;  F_N(z) = u+O(t), \qquad  Im \; F_N(z) = 2t\sqrt{1-u^2} + o(t)
$$
so $F_N(\Xi)\subset \Xi$.

Now we compute, for $z\in\Xi$,
$$
  F_N'(z) = \frac{t}{N} \sum_j\frac{1}{(z-y_j)^2} 
  = F'(z) + O(N^{-\lambda/4}) 
$$
(here we used \eqref{ellgood} with $\ell=2$ and observed
that $F_N' = tf_N''$),
and
$$
  F'(z) = -2t\Big[ 1 - \frac{z}{\sqrt{z^2-1}}\Big]
$$
with $F'(z)= O(t)$ if $z\in \Xi$.
Thus $|F_N'(z)|\le 1/2$ for $z\in \Xi$, so $F_N$ is
a contraction on $\Xi$ and thus
 \eqref{qNeq}
has a unique solution, which is $q_N$.

Comparing the two solutions, we have
$$
   |q_N-q| = |F_N(q_N)- F(q)| \le \sup_{z\in\Xi} |F_N'(z)|
 |q_N-q| + |F_N(q) - F(q)|.
$$
Since $y\in \cY$, we get
$$
  |F_N(q) - F(q)| \le t \Bigg| \frac{1}{N}\sum \frac{1}{q_N-y_j} 
   -  \int 
  \frac{\varrho_{sc}(y)\rd y}{z-y}\Bigg| = \frac{Ct}{N^{\lambda/4}}
$$
thus
\be
   |q_N-q| \le \frac{Ct}{ N^{\lambda/4}}.
\label{qqN}
\ee

\subsection{Evaluating the integrals}

Using Laplace asymptotics, we compute the integrals in \eqref{repr}.
We choose the horizontal curves $\gamma_\pm$ to pass through
the two saddles $q^\pm= a\pm ib$ of $f$ (see \eqref{qsol}),
 i.e. we set $\om = b$ (see the definition of $\gamma^\pm$ after
\eqref{ck}). The vertical line $\Gamma$ is shifted to pass through
the saddles, i.e. $Re\; \Gamma = a$. Moreover, 
if necessary, we deform $\Gamma$
in a $O(N^{-1})$-neighborhood of $a$ 
 so that $\min_j \mbox{dist}(\Gamma, y_j) \ge N^{-2}$
and  $\mbox{dist}(\Gamma, a_N) \ge N^{-2}$;
this is always possible.

We split the integrals as follows
$$
    \frac{1}{N\varrho}\cK_N(u,u+\frac{\tau}{N\varrho}; y) =
    A^{++}+A^{+-}+A^{-+}+A^{--}
$$
according to whether  $Im \; z$ and $Im\; w$ are positive or negative, e.g.
\be
   A^{\pm \pm}:= N \int_{\gamma^\pm}
  \frac{\rd z}{2\pi i}\int_{\Gamma^\pm} \frac{\rd w}{2\pi i}
  h_N(w) g_N(z,w) e^{N(f_N(w)-f_N(z))}
\label{A+++}
\ee
where 
 $\Gamma_+=\Gamma \cap \{ w\; : \; Im \, w\ge0\}$ and 
$\Gamma_-=\Gamma \cap \{ w\; : \; Im \, w\le0\}$. We will
work on $A^{++}$, the other three integrals are treated similarly.

The main contribution to the integral $A^{++}$
will come from an $\e$-neighborhood in $z$ and $w$
of the saddle point $q_N=q_N^+$.
 The radius $\e$ will be chosen such that after
a local change of variable $f$ and
$f_N$ become quadratic near the saddle. We now
explain the local change of variable.

Since $f(z):\bC\to\bC$ is
an analytic function with $f'(q)=0$ and $f''(q)\ne 0$
for $q=q^+$,
there exists an invertible analytic map $\phi: z\to \phi(z)$
in 
$$
    D_\e: = \{ z\; : \; |z-q|\le\e\}
$$
with $\phi(q)=0$, $\phi'(q)=\sqrt{tf''(q)}$ such that
\be
      f(z) = f(q)+ \frac{1}{2t} [\phi(z)]^2   \qquad z\in D_\e
\label{morse}
\ee
with
\be
    \phi(z) = \sqrt{tf''(q)}(z-q)(1+ O(z-q)), \qquad z\in D_\e.
\label{phi}
\ee
Here $\e$ must satisfy
\be
  \e \le \frac{|f''(q)|}{2\sup_{D_\e} |f'''(z)|}
\label{eps}
\ee
 we also assume that $\e \le \eta$.
We will choose $\e=ct$ with a small $c$, depending on $u$.
We have 
\be
 f''(q) = t^{-1} + O(1), \qquad
\sup_{D_\e} |f'''(z)| \le C 
\label{fder}
\ee
 from
the explicit formula \eqref{fprime}, so \eqref{eps} 
is satisfied. Note that $\phi'(q)= \sqrt{tf''(q)} = 1+ O(t)$.

We have a similar change of variables for $f_N$,
i.e. $\phi_N$ with the properties that 
\be
\phi_N(q_N)=0, \qquad \phi'_N(q_N)
= \sqrt{tf''_N(q_N)} = 1 + O(t)
\label{phider}
\ee
 and
\be
      f_N(z) = f_N(q_N)+ \frac{1}{2t}[\phi_N(z)]^2   \qquad z\in D_{\e,N}
   = \{ z\; : \; |z-q_N|\le \e\}
\label{morseN}
\ee
with
\be
    \phi_N(z) = \sqrt{tf''_N(q_N)}(z-q_N)(1+ O(z-q_N)), \qquad z\in D_{\e,N}.
\label{phiN}
\ee
This holds if
$$
   \e \le \frac{c |f''_N(q_N)|}{\sup_{D_{\e,N}} |f'''_N(z)|}.
$$
For $y\in \cY$, we
 have $f''_N(q_N)  = t^{-1}\big[1+O(N^{-\lambda/4})\big]$ and $|f'''_N(z)|\le 
Ct^{-2}N^{-\lambda/4}$ by \eqref{ellgood} and \eqref{fder},
thus we can choose $\e = ct$ for some small constant $c\le \sqrt{1-u^2}$.

Moreover we have $|\phi_N(z)|\le C|z-q_N|$ for $|z-q|\le ct$, so
by Cauchy formula $|\phi'_N(z)| \le C$ and $|\phi''_N(z)|\le Ct^{-1}$
for  $|z-q|\le ct$ (maybe after reducing $c$). The same formulas hold for
$\phi$ as well. 
We also have
$$
  | \phi'(q) - \phi'_N(q_N)| \le \Big| \sqrt{t f''(q)} - \sqrt{t f_N''(q)}\Big|
  +  \Big| \sqrt{t f_N''(q)} - \sqrt{t f_N''(q_N)}\Big|
  \le CN^{-\lambda/4},
$$
where in the first term we used  \eqref{ellgood} and in the second
we used $|f'''_N(z)|\le Ct^{-2}$.

{F}rom \eqref{phi} and \eqref{phiN} 
 we have 
\be
    |\phi(z)- \phi_N(z) |\le  | \phi'(q) - \phi'(q_N)||z-q|
  + |\phi'(q_N)||q-q_N| + C|z-q|^2 \le  Ct N^{-\lambda/4}
\label{phip1}
\ee
and then by contour integration
\be
    |\phi'(z) - \phi_N'(z)|\le  Ct N^{-\lambda/4}
\label{phip}
\ee
for any $z$ with $|z-q|\le ct$. Therefore the maps $\phi$ and $\phi_N$
are $C^1$-close within $D_\e$ and both of them are $C^1$-close
to the shift map $z\to z-q$.

\begin{figure}
\begin{center}
\epsfig{file=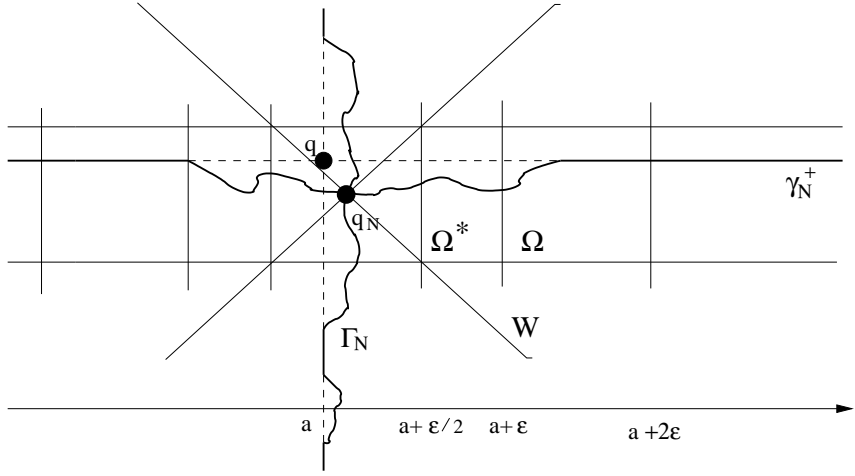,scale=.75}
\end{center}
\caption{Integration contours
around the saddle $q_N=q_N^+$}\label{fig:saddle}
\end{figure}

\bigskip

We first consider  the $z$ integration. Recall that $q_N=q^+_N=a_N+ib_N$
from \eqref{qsol}. 
We fix a small positive constant $c_1\ll 1$ and we define the domains
$$
  \Omega: = \Big\{ z=x+iy\; : \; |x-a_N|\ge \e, |y-b_N|\le c_1\e/2\Big\}
$$
$$
  \Omega^*: = \Big\{ z=x+iy\; : \; |x-a_N|\ge \e/2, |y-b_N|\le c_1\e/2\Big\}
$$
and
$$
   W: = \Big\{ z= x+iy \; : \; |x-a_N|\le 2\e\; , \; |y-b_N|\le c_1 |x-a_N| 
\Big\}
$$ 
where $\e=ct$.
Recall that $\gamma^+$ was the horizontal line going through 
$q=a+ib$, the saddle of $f$.
We will deform
$\gamma^+$ to $\gamma_N^+$ so that it passes through $q_N$
and it matches with $\gamma^+$ at the points $a_N\pm 2\e + ib$. 
Within the regime $|Re \; z -a_N|\le \e$, we define $\gamma_N^+$
by the requirement that $Im \; \phi_N =0$ along $\gamma_N^+$.
Since $\phi_N(z)$ is close to the map $z\to z-q_N$ by \eqref{phiN},
clearly $\gamma_N^+$ is almost horizontal curve in small neighborhood
of $q_N$, so it remains in $W$ until it reaches the vertical lines
$|Re\; z-a_N|= \e$. In the regime $\e\le|Re \;z -a_N |\le 2\e$,
we require that $\gamma_N^+$ matches with 
 $\gamma^+$ at the points $a_N\pm 2\e + ib$
and it remains in the wedge $W$. In the outside regime, $|Re \; z- a_N|\ge 2\e$
we set $\gamma_N^+=\gamma^+$, in particular $\gamma_N^+ \subset W\cup \Omega$
(see Fig. \ref{fig:saddle}).

\begin{lemma}\label{lm:z}
We have
\be
    Re \big[f_N(x+iy)-f_N(q_N)\big]
 \ge \frac{1}{12 t}(x-a)^2 \qquad \mbox{for}\;\;  x+iy\in \Omega
\label{reflower}
\ee
and
\be
    Re \big[f_N(z)-f_N(q_N)\big]\ge 0 \qquad \mbox{for}
 \quad z\in W \; \mbox{and}\quad |Re\; z -a|\le \e
\label{reflowerin}
\ee
\end{lemma}

{\it Proof.} The second statement \eqref{reflowerin} follows from
the normal form \eqref{morseN} and
the fact that for $z \in W$ we have $|Im \; (z-q_N)| \le c_1 |Re (z-q_N)|$,
i.e. $Re (z-q_N)^2 \ge 0$, and $\phi_N$ is close to the map $z\to z-q_N$
in $W$, so $Re [\phi_N(z)]^2 \ge0$ for $z\in W$.

For the first statement, we assume $x\ge a$, the 
case $x\le a$ is analogous. We get 
by explicit calculation 
$$
   Re \; f'(x+iy) \ge \frac{1}{2t}(x-a), \quad
 \mbox{for}\;\; x+iy \in\Omega^*, \; x\ge a.
$$
Using \eqref{ellgood} for $\ell =1$, we have
\be
    Re\; \partial_x f_N(x+iy) \ge
   Re \; f'(x+iy) - CN^{-\lambda/4}
  \ge \frac{1}{3t}(x-a)  \quad
 \mbox{for}\;\; x+iy \in\Omega^*\; x\ge a
\label{fx}
\ee
(the error is absorbed since $|x-a|\ge ct/2$ for  $x+iy \in\Omega^*$).
Since $Re [f_N(z) - f_N(q_N)]\ge 0$ on the vertical lines $|x-a|=\e/2$,
$|y-b|\le c_1\e/2$, we can integrate the inequality \eqref{fx} to
obtain \eqref{reflower}. \qed

\bigskip

In order to estimate the $w$ integration along 
 $\Gamma^+$ parametrized as $a+is$, $s\ge 0$,
we analyze the behaviour of $Re \, f$ along $\Gamma^+$.
For $|x-a|\le Ct$ and $y\in \bR$ we first compute 
$$
     Re\; \partial_y f_N(x+iy) = - Im\; f'_N(x+iy) = - Im \; f'(x+iy)
    + O( N^{-\lambda/4})
$$
which holds for $|y|\ge \eta$.  
By explicit computation,  and using $f'(a+ib)=0$, 
$$
  - Im \; f'(x+iy) 
 = - (y-b)\Big( \frac{1}{t} + 2 \Big) +O(t) +O(y^2)
$$
if $|y|\le \frac{1}{2}\sqrt{1-u^2}$, $|x-a|\le Ct$ for some large $C$.
Thus we have
$$
     Re\; \partial_y f_N(x+iy) \le -\frac{y-b}{3t}, \qquad  \eta\le |y|\le
  \frac{1}{2} \sqrt{1-u^2}\;\;\mbox{and} \; \; y-b\ge \e/2,
$$
where $\e=ct$ with a small $c$ as before and a similar lower
bound holds for $y-b\le -\e/2$.
Defining 
$$
   \wt\Omega:= \Big\{ w=x+iy\; : \; \e\le |y-b_N|,\; 
\eta\le y\le  \frac{1}{2} \sqrt{1-u^2},\;\;
 \; |x-a_N|\le c_1\e/2\Big\}
$$
$$
  \wt W: = \Big\{ w=x+iy\; : \; |y-b_N|\le 2\e, \; |x-a_N|\le c_1|y-b_N|\Big\}
$$
analogously to $W$ before, we easily obtain
\be
    Re \,\big[ f_N(x+iy)-  f_N(q_N)\big] 
 \le -\frac{1}{18t}(y-b)^2 \quad \mbox{for}
   \;\; x+iy\in \wt \Omega 
\label{wlow}
\ee
and
\be
    Re \,\big[ f_N(w)-  f_N(q_N) \big] \le 0
   \;\; w\in \wt W, \; \mbox{and} \; |Im\, w-b|\le\e,
\label{wlowin}
\ee
similarly to the proof of Lemma \ref{lm:z}.

The regimes  $0\le y\le\eta$ and $y\ge\frac{1}{2} \sqrt{1-u^2}$
 are treated directly.
We use
\be
\begin{split}
  Re \partial_y f_N (x+ i y)  =  &- Im  
\left [ t^{-1}(z-u) +\frac{1}{N}\sum_j\frac{1}{z-y_j} \right]   \\
= &   y  \left [ -t^{-1}  + \frac{1}{N}\sum_{j }
 \frac{1}{(x-y_j)^2 + y^2} \right ]
\ge - y /t.
\label{3.1}
\end{split}
\ee
Hence for $0\le y \le \eta$ we have
\[
Re\; \big[ f_N (x+iy)-f_N(q_N)\big] \le Re\; \big[ f_N (x+i\eta)-f_N(q_N)\big]
 +\frac{\eta^2}{2t} \le -\frac{1}{36t}(y-b)^2
\]
from \eqref{wlow}, if $\eta_0$ is sufficiently small, see \eqref{eta}.

If $y \ge\frac{1}{2} \sqrt{1-u^2}$, then
\[
   \frac{1}{N}\sum_{j } \frac{1}{(x-y_j)^2 + y^2} \le\frac{4}{\sqrt{1-u^2}},
\]
hence 
\be
  Re \; \partial_y f_N (x+ i y)  \le - y /2t
\label{3.3}
\ee
and thus $Re \; f_N (x+ i y)\le  -y^2/4t$ in this regime.
Summarizing these results, we have
\be
Re\; \big[ f_N (x+iy)-f_N(q_N)\big] 
 \le -\frac{1}{36t}(y-b)^2 
\label{wlow1}
\ee
holds for any $y\in \bR$ and $|x-a|\le c_1\e/2$.

\bigskip 

We can define a new contour $\Gamma_N^+$ similar to the $\gamma_N^+$.
It follows the path where $\phi_N$ has  
zero imaginary part when $|Im\; w -b|\le \e/2$  and 
then  it returns to $\Gamma^+$  when $|Im\; w -b|\ge \e$. 
We recall that 
 $\min_j \mbox{dist}(\Gamma_N^+, y_j) \ge N^{-2}$
and $\mbox{dist}(\Gamma, a_N) \ge N^{-2}$
by the choice of $\Gamma$.

With the paths $\gamma_N^+$ and $\Gamma_N^+$ defined, we can now
do the integration
\be
  A^{++}:
  = N \int_{\gamma_N^+} \frac{\rd z}{2\pi i}\int_{\Gamma_N^+}
  \frac{\rd w}{2\pi i}
  h_N(w) g_N(z,w) e^{N(f_N(w)-f_N(z))}.
\label{A}
\ee
Near the saddle 
we  need the  bounds 
\be
 |g_N(z,w)|\le C/t, \quad
 |\partial_z g_N(z,w)| \le C/t^2 , \quad
 |h_N(w)|\le C 
\label{gh}
\ee 
if $|z-(a+ib)|\le \e$, $|w-(a+ib)|\le\e$. 
In order to make sure that these bounds are satisfied, we fix the constant 
$r= \text{Re } q_N (u_*)$ in (\ref{repr}). Here $q_N (u^*)$ is the unique 
solution with positive imaginary part of the saddle point equation 
(\ref{fNroot}), with $u$ (which is actually a short hand notation
 for $u^{(N)}$) replaced by the fixed $u_*$. Note that, since 
$|u^{(N)}-u_*|\leq C/N$, we find that the real part of the exponent
of  $h_N(w)$ (see (\ref{def:hN})) is bounded,
$|r - \text{Re} w|/t\rho \leq C$, as $w$ runs through $\Gamma$.

This choice also guarantees that, away from the saddle, 
\be
  |h_N(w)|\le C e^{Ct^{-1}|Re\, w- a|}, \quad |g_N(z,w)| \;  \le CN^3
\label{gh1}
\ee
that hold for $|Im\; z|\ge \eta$, $Im \; w\ge 0$. These bound
follow from \eqref{def:gN}, \eqref{def:hN} and \eqref{id} and
when $w$ is near the real axis, we also used that $\Gamma_N$
is away from the $y_j$'s.

\bigskip

The  integration in $A^{++}$ (see \eqref{A}) will be  divided into regimes
near the saddle $q_N$ (``inside'') or away from the saddle (``outside''):
\be
    A^{++} = A_{ii} + A_{io}+ A_{oi} + A_{oo}.
\label{A++}
\ee
Recall that $|q_N-q| =o(t)$ and $q= q^+= a+ ib$ (see \eqref{qsol}).
For example 
$$
A_{io}:= N \int_{\gamma^+_N}   \chi( Re \; z- a) \frac{\rd z}{2\pi i} 
   \int_{\Gamma^+_N} (1- \chi( Im \; w - b))  \frac{\rd w}{2\pi i}
  h_N(w) g_N(z,w) e^{N(f_N(w)-f_N(z))} ,
$$  
where $\chi $ is the characteristic function of the interval
$[-\e,\e]$. The other $A$'s are defined analogously.

Using \eqref{reflower} \eqref{reflowerin}
and \eqref{wlow1}, we have
$$
|A_{io}| \le N\int_{\gamma_N} \,\rd z\, 
  \chi( Re \; z- a)
\int_{\Gamma_N} (1- \chi( Im \; w - b)) |g_N(z,w)||h_N(w)|\;
  e^{N Re\; [f_N(w)- f_N(q_N)]} \rd w.
$$
The integral of the exponential term is bounded by 
$$
 \int_{|y-b|\ge \e=ct}  \rd y \; e^{-cN(y-b)^2/t}
  \le e^{-cNt}.
$$
Taking into account \eqref{gh} and \eqref{gh1}, we see
that $|A_{io}| \le e^{-cNt}$ since $t= N^{-1+\lambda}$.
 Similarly we 
can bound all other terms with an outside part. 
When $|Re\, z -a|\ge ct\gg N^{-1}$, then the exponential growth
of $h_N$ in \eqref{gh1} will be controlled by the
Gaussian decay of 
$$
   e^{ -N \, Re[ f_N(z)-f_N(q_N)]}\le e^{-cNt^{-1}|Re \, z-a|^2}
$$
from \eqref{reflower}.

Finally, we have to compute the contribution of the saddle, i.e.
the term $A_{ii}$. We
let $\wt\gamma$ be the part of $\gamma_N^+$ with $|Re\; \gamma_N-a|\le \e$
and similarly defined  $\wt \Gamma$. 
Recall that $Im \; \phi_N =0$ on $\wt\gamma$. 
{F}rom standard Laplace asymptotics calculation, we have
$$
    \int_{\wt\gamma}  e^{-N[f_N(z)-f_N(q_N)]} h_N(w)g_N(z,w) \rd z
  =  \int_{\wt\gamma}  e^{-N[\phi_N(z)]^2/2t} 
h_N(w)g_N(z,w) \rd z
$$
\be
 = 
 \sqrt{\frac{2\pi}{N f_N''(q_N)}}\Bigg[   
  h_N(w)g_N(q_N,w) + \Omega(w)\Bigg]
\label{statph}
\ee
using \eqref{phider} with
$$
  |\Omega(w)|\le C \sqrt{\frac{t}{N}} \max_{z\in D_\e}  |\partial_z g_N(z,w)||h_N(w)|
$$
Using \eqref{gh}, we have
$$
  |\Omega|\le  Ct^{-2}\sqrt{\frac{t}{N}} = \frac{C}{t} \frac{1}{\sqrt{Nt}}
$$
while the main term  in the bracket on the r.h.s. of
\eqref{statph} is of order $t^{-1}$. Analogously performing the
$\rd w$ integration, we obtain that 
$$
A_{ii} =  \frac{-1}{2\pi f_N''(q_N)} g_N(q_N, q_N) h_N ( q_N)
\Big[ 1+ O\Big(\frac{1}{\sqrt{Nt}}\Big)\Big] = 
 \frac{-h_N (q_N)}{2\pi} \Big[ 1+ O\Big(\frac{1}{\sqrt{Nt}}\Big)\Big],
$$
where we also used $ g_N(q_N, q_N) = f_N''(q_N)$ following from \eqref{id}.
So far we considered the saddle $q_N=q_N^+$ with positive
imaginary part for both the $z$ and $w$ integrals.
The same calculation can be performed at the saddle $z=w=q_N^-$.
The mixed case, when $z$ is integrated near one of the saddles
and $w$ is near the other one, gives zero contribution, since
$g_N(q_N^-, q_N^+)= g_N(q_N^+, q_N^-) =0$ by \eqref{id}.
Adding up the contributions of the two relevant saddles,
 $z=w=q_N^+$ and $z=w=q_N^{-}$, taking into account
the opposite orientations of the two pieces of $\gamma_N$,
one obtains
$$
    \frac{1}{2 \pi} \Big[  - h_N (q_N^+) +
   h_N ( q_N^-)\Big] =
\frac{1}{2\pi\tau} \Big( - 
  e^{-\tau (q_N^+-r)/t\varrho}+ 
   e^{-\tau (q_N^- -r)/t\varrho} \Big) = \frac{\sin \pi\tau}{\pi\tau}(1+o(1)),
$$
 where we used the choice $r = \text{Re } q^{\pm}_N (u^*)$ (see after (\ref{gh})), which guarantees that $|r-\text{Re} q_N^\pm| \to 0$ as $N\to\infty$, and the equations  \eqref{not},
\eqref{qsol}, and \eqref{qqN}. This completes the proof of Proposition
\ref{prop:local}.
\qed

\section{Proof of the main theorems}\label{sec:mainthm}

{\it Proof of Theorem \ref{mainthm}.}  We follow the
notations of Proposition \ref{meascomp}.
In Proposition \ref{sinjoh}  we have shown that the
sine kernel holds for the measure $e^{t\cL}G_t$ 
if $t=N^{-1+\lambda}$. More precisely, let $p_{N,t}(x)$,
 denote the density function of the eigenvalues $x=(x_1, \ldots, x_N)$
w.r.t. $e^{t\cL}G_t$ and let
 $p_{N,t}^{(2)}$ be the two point correlation function,
defined analogously to \eqref{corrfn}.
Similarly, we define $p_{N,c}(x)$ and  $p_{N,c}^{(2)}$
for the eigenvalue density and 
two point correlation function 
w.r.t. truncated measure $F_c=v^{\otimes n}$.

In Proposition
\ref{sinjoh} we  showed that
\be
 \lim_{N\to\infty}\int_{\bR^2} \frac{1}{\varrho^2} p_{N,t}^{(2)}
\Big( u+ \frac{\al}{N\varrho}, u+\frac{\beta}{N\varrho}\Big) O(\al,\beta)
 \rd \al\rd \beta =  \int_{\bR^2} O(\al,\beta) 
\Big[1-\Big(\frac{\sin \pi(\al-\beta)}{\pi(\al-\beta)}\Big)^2\Big]
 \rd \al \rd \beta.
\label{sint}
\ee
for any $|u|<2$ and with the notation $\varrho=\varrho_{sc}(u)$.
(We remark that $p_{N,t}^{(2)}$ was  denoted by $\wt p_N^{(2)}$
in Proposition \ref{sinjoh}
and the condition $|u|<2$ is translated into $|u|<1$
after rescaling.)

To prove \eqref{maineq}, we thus only need to control
the difference as follows
\bigskip
$$ 
 \Bigg|
\int \Big[ p^{(2)}_N\Big( u+ \frac{\al}{N\varrho}, u+\frac{\beta}{N\varrho}
  \Big)
 -p^{(2)}_{N,t}\Big( u+ \frac{\al}{N\varrho},
   u+\frac{\beta}{N\varrho}\Big)\Big]O(\al,
\beta) \rd \al\rd \beta
 \Bigg| \le (I)+ (II),
$$
where
$$
  (I): =  \Bigg|
\int \Big[ p^{(2)}_N\Big( u+ \frac{\al}{N\varrho}, u+\frac{\beta}{N\varrho}
  \Big)
 -p^{(2)}_{N,c}\Big( u+ \frac{\al}{N\varrho},
   u+\frac{\beta}{N\varrho}\Big)\Big]O(\al,
\beta) \rd \al\rd \beta
 \Bigg|,
$$
$$
  (II): =  
\int \Big| p^{(2)}_{N,c}\Big( u+ \frac{\al}{N\varrho}, u+\frac{\beta}{N\varrho}
  \Big)
 -p^{(2)}_{N,t}\Big( u+ \frac{\al}{N\varrho},
   u+\frac{\beta}{N\varrho}\Big)\Big|\, |O(\al,
\beta)| \rd \al\rd \beta.
$$
Using \eqref{FFtilde}, we have
$$
  (I)\le N^2 \|O\|_\infty \int |F-F_c|\rd\mu^{\otimes n} \le Ce^{-cN^c}
\to 0
$$
with some $c>0$ as $N\to\infty$.
To estimate $(II)$, we have
\be
\begin{split}
(II)  & \le \int \Bigg| \frac{p^{(2)}_{N,c} }{p_{N,t}^{(2)}}
\Big( u+ \frac{\al}{N\varrho}, u+\frac{\beta}{N\varrho}\Big)-1\Bigg|
   p_{N,t}^{(2)}
\Big( u+ \frac{\al}{N\varrho}, u+\frac{\beta}{N\varrho}\Big)
  |O(\al, \beta)| \rd \al\rd \beta
\\
  & \le  \Big( \int  \Big[ \frac{p_{N,c}^{(2)}}{p_{N,t}^{(2)}}
\Big( u+ \frac{\al}{N\varrho}, u+\frac{\beta}{N\varrho}\Big)-1\Big]^2  
p_{N,t}^{(2)}
\Big( u+ \frac{\al}{N\varrho}, u+\frac{\beta}{N\varrho}\Big) |O(\al, \beta)| 
\rd \al\rd \beta \Big]^{1/2} \\
& \quad\times
\Big[ \int p_{N,t}^{(2)}
\Big( u+ \frac{\al}{N\varrho}, u+\frac{\beta}{N\varrho}\Big) |O(\al,\beta)|
 \rd \al\rd \beta \Big]^{1/2}.
\label{long}
\end{split}
\ee
Using \eqref{sint} for the observable $|O|$ instead of $O$,
the second factor on the r.h.s. of \eqref{long}
is bounded. Since $O$ is bounded,
the first factor is smaller than
\be
\begin{split}\label{NN}
  C   \Bigg[N^2\varrho^2 \int  
  \Big[ \frac{p^{(2)}_{N,c}(z,y)}{p_{N,t}^{(2)}(z,y)}
  -1\Big]^2
   p_{N,t}^{(2)}(z,y) \rd z\rd y \Bigg]^{1/2} & 
\le  C \Bigg[N^2\varrho^2 \int  
  \Big( \frac{p_{N,c}(x)}{p_{N,t}(x)}
  -1\Big)^2
   p_{N,t}(x)\rd x\Bigg]^{1/2}
\\
 & \le C \Bigg[N^2\varrho^2 
 \int  \frac{\big|e^{t\cL}G_t- F_c\big|^2}{e^{t\cL}G_t} 
\rd \mu^{\otimes n} \Bigg]^{1/2} \\
& \le C N^{-1+4\lambda}.
\end{split}
\ee
Here in the first step we used that
 the quantity $D(f,g)= \int |f/g-1|^2g$
for two probability measures $f$ and $g$ decreases when
taking marginals. In the second step, we used that
$D(f,g)$ decreases when passing the probability
laws from matrix elements to the induced
probability laws for the eigenvalues.
Finally, we used  the estimate
\eqref{FF}.
This completes the proof of Theorem \ref{mainthm}. \qed

\bigskip

{\it Proof of Theorem \ref{mainthm2}.}
We first prove Theorem \ref{mainthm2} for the ensemble $\wh H + aV$ 
with $a=N^{-1/2+\lambda/2}$
(see the beginning of Section \ref{sec:timeevolved} for the necessary 
rescaling).
Let $\EE$ denote the expectation with respect to this
ensemble and let $\EE_y$ denote the expectation with respect
to the density $x\to q_S(x,y)$ for any fixed $y$ and $S= a^2/N=N^{-2+\lambda}$.
Then we have
\be
  \E \, \Lambda (u; s, \cdot ) 
   = \int  \EE_y \; \Lambda (u; s, \cdot ) {\bf 1}(y\in\cY) \rd \wh \PP(y) +
 \int  \EE_y \; \Lambda (u; s, \cdot ) {\bf 1}(y\in\cY^c) \rd \wh \PP(y)
\label{lambdasplit}
\ee
by recalling \eqref{def:qs}.
The second term can be estimated by using $|\Lambda|\le N$ and \eqref{good} as
\be
   \int  \EE_y \; \Lambda (u; s, \cdot ) {\bf 1}(y\in\cY^c) \rd \wh \PP(y)
  \le C Ne^{-cN^{\lambda/4}}.
\label{tail}
\ee 
For the first term in \eqref{lambdasplit},
 we use the exclusion-inclusion principle to compute
\be
\begin{split}\label{8.1}
\E_y \, \Lambda (u; s, \cdot ) = 
 \frac 1 { 2 N t_N \varrho} \sum_{m=2}^N (-1)^m   & 
\int_{-t_N }^{t_N }   \rd v_1 \ldots  \int_{-t_N }^{t_N}  
 \rd v_m  {\bf 1} \Big\{ \max |v_i-v_j| \le \frac s { N \varrho}\Big\} \\ 
&   \times \,   {N \choose m} \,  \wt p^{(m)}_{N,y,S}(u+v_1 , u+v_2,
 \ldots, u+ v_m)  
\end{split}
\ee
with $\varrho=\varrho(u)$ (see \eqref{def:varrho}) and recall that
$\wt p^{(m)}_{N,y,S}$ denote the correlation functions of $q_S(x,y)$
(see \eqref{tildecorr}). After a change of variables,
\be
\begin{split} 
\E_y \; \Lambda (u; s, \cdot )= \; & \; \frac 1 { 2 N t_N \varrho}
 \sum_{m=2}^\infty (-1)^m   
\int_{-N\varrho t_N }^{N\varrho t_N  }   \rd z_1 \ldots 
 \int_{-N\varrho t_N }^{N\varrho t_N  }
  \rd z_m \\ 
&   \times\,   {N \choose m} \,  \frac{1}{(N\varrho)^m}
  \wt p^{(m)}_{N,y,S} \Big(u+ \frac {z_1}{N \rho}  , 
 \ldots, u+ \frac {z_m}{N \rho}\Big) {\bf 1} \Big\{\max |z_i-z_j| 
\le  s\Big\}   \\
= \; & \;
\frac 1 { 2 N t_N \varrho} \sum_{m=2}^\infty (-1)^m  m 
\int_{-N\varrho t_N }^{N\varrho t_N  }   
\rd z_1  \int_0^s \rd a_2 \ldots \int_0^s \rd a_m  \\
&   \times\,   {N \choose m} \,  \frac{1}{(N\varrho)^m}\;
  \wt p^{(m)}_{N,y,S} \Big(u+ \frac {z_1}{N \rho}  , 
u+ \frac {z_1+a_2}{N \rho} ,
 \ldots, u+ \frac {z_1+a_m}{N \rho}\Big),
\end{split}
\ee
where the factor $m$ comes from considering the integration
 sector $z_1\le z_j$,
$j\ge 2$.
Taking $N\to \infty$ and using Proposition \ref{prop:local}, we get
\be
   \lim_{N\to\infty} \E_y \; \Lambda (u; s, \cdot ) 
= \sum_{m=2}^\infty \frac {(-1)^m }{(m-1)!}  
  \int_0^s \rd a_2 \ldots  \int_0^s  \rd a_m   \,   \,  \det 
 \left ( \frac {\sin \pi(a_i-a_j)} {\pi(a_i-a_j)}  \right )_{i,j=1}^m,
\label{fred}
\ee
where in the last determinant term we set $a_1=0$. 
The interchange of the limit and the summation can be
justified by noting that the exclusion-inclusion principle guarantees
that \eqref{8.1} is an alternating series where the difference between
the sum and its $M$-term truncation can be controlled by the $(M+1)$-th term
for any $M$. 
We note that
the left hand side of \eqref{fred} is $\int_0^s p(\al)\rd \al$,
where $p(\al)$ is the second derivative of the Fredholm determinant
$\det (1-\cK_\al)$  (see \eqref{maineq2}).
Combining \eqref{fred}
with the estimate \eqref{tail}, we have
\be
\label{elambda}
\lim_{N\to\infty} \E \; \Lambda (u; s, \cdot ) =\int_0^s p(\al)\rd \al.
\ee
After rescaling \eqref{resc}, we also conclude that the limit
of the expectation of $\Lambda$ with respect to the
time evolved  ensemble $e^{t\cL}G_t$ (see Proposition \ref{meascomp})
is given by right hand side of \eqref{elambda}.

Finally, the difference of  the expectation of $\Lambda$ with respect to the
measure 
 $e^{t\cL}G_t$ and  w.r.t. the initial ensemble $F$ vanishes since
 $|\Lambda|\le N$
and $\mbox{Var}(e^{t\cL}G_t, F) \le CN^{-2+4\lambda}$ (see \eqref{FFtilde}
and \eqref{FF}).
This completes the proof of Theorem \ref{mainthm2}. \qed

\section{Some extensions and comments}\label{sec:relax}

In this section we explain how to relax some of the conditions
on the initial distribution $\nu$.

We first  explain how to extend our proof
to include distributions $\nu$ with compact support. Take for example
a density w.r.t. the Gaussian measure $\rd \mu(x) = e^{-x^2}$
that is given by a nice bump function $u(x)$ 
supported in $[-1, 1]$ decaying like $(1\pm x)^m$ near the boundary $x=\pm 1$.
Clearly, for any $m$ fixed this distribution violates the assumptions  of
  Theorem \ref{mainthm}. 
We now show that for $m$ large enough, it is still possible to
 prove the universality. 
Define a new distribution with density 
\be
q (x) =  \frac{\tau^m + u(x)}{1+ \tau^m}
\label{q}
\ee
with a small parameter $\tau>0$ to be determined later.
Near the edge $1$ we have $L q(1-y) \lesssim C  y^{m-2}$ for $0\le y \ll 1$
with some $m$-dependent constant $C$. 
We thus need the condition 
\[
 C y^{m-2} \le  t^{-1}  [\tau^m + y^m ] ,\qquad 0\le y \ll 1,
\]
to guarantee that $(1- t L) q$ 
is a probability density. 
 This inequality holds  if
\[
\tau^2 \ge C t.
\]
The other conditions 
concerning $L^2$ and $L^3$ (see \eqref{AB}) can 
be handled similarly. Choosing $\tau= Ct^{1/2}$, the total variation norm is
bounded by 
\[
\int |q^{\otimes n} - u^{\otimes n} |  \rd  \mu^{\otimes n}  
\le Cn \tau^{m} = C n t^{m/2}.
\]
Since $n= N^2$ and $t= N^{-1+ \e}$, we have 
\[
\int |q^{\otimes n} - u^{\otimes n} | \rd  \mu^{\otimes n}  \le
 C_m N^{2- m/2+m \e /2}.
\]
Let, say,  $m\ge 9$, then the error term will be smaller 
than $N^{-2-\delta}$ with some $\delta>0$ and this will imply 
Theorem \ref{mainthm} for the initial distribution $u$.
The modification of $u$ in \eqref{q}
 can certainly be more sophisticated to reduce the exponent $m$.

\bigskip 

Second, we show that the Gaussian  decay condition \eqref{cond2}
can be replaced 
by the exponential decay \eqref{cond2relax}.
For any $\ell > 0$ define 
\[
\nu_\ell (x) =    \nu (x+ a_\ell ) {\bf 1}(|x| \le \ell) /Z_\ell,
\]
where $a_\ell$ and $Z_\ell$  are chosen so that 
\[
\int x  \;\rd \nu_\ell  = 0, \qquad  \int \rd \nu_\ell (x) = 1.
\]
Due to the assumption \eqref{cond2relax}, we have 
\[
|a_\ell| + |Z_\ell-1| \le e^{ - c \ell}.
\]
Let $ \wt \nu_\ell (x) = \ell  \nu_\ell (x \ell)$. Clearly, the
 random variable $x$ distributed according to $  \tilde  \nu_\ell$ is bounded
by $1$, in particular it has a finite Gaussian moment.
  Denote the variance of $ \tilde   \nu_\ell$ by $\sigma_\ell^2$
and we have $\sigma_\ell =1/\ell + O(e^{ - c \ell})$. We will neglect all the 
exponential small terms $O(e^{ - c \ell})$ and assume  $\sigma_\ell =1/\ell$.
Similar cutoff and rescaling applies to the distribution of the diagonal elements.

Consider the random matrix generated by the measure $\nu_\ell$ and 
$  \tilde  \nu_\ell$  and denote the probability law of the eigenvalues 
by $f_\ell$ and $  \tilde  f_\ell$. Since all quantities introduced 
below can be defined w.r.t. to both $\nu_\ell$ and $  \tilde   \nu_\ell$, 
we will only give explicit definitions 
for $\nu_\ell$. 
Recall  the Stieltjes transform of the eigenvalue distribution w.r.t.
 $ \nu_\ell$  is defined as  
\be
 m_\ell= m_\ell(z) = \int_\bR \frac{\rd F_\ell (E)}{E-z}\,,
\label{Sti}
\ee
where $F_\ell $ is the empirical distribution function of the eigenvalues. 
Then the empirical density of eigenvalues and 
$ \tilde m_\ell$ converges to the rescaled semicircle law
\[
 \wt\rho^\ell_{sc} (x) =  \ell   \rho_{sc} (x  \ell)   = 
 \frac \ell  {2 \pi } \sqrt {4 - x^2\ell^2},\quad 
 \wt m_{sc}^\ell (z) = \ell  m_{sc} (z \ell).
\]

We now follow the proof given in \cite{ESY3} to prove the local 
semicircle law Theorem 4.1 \cite{ESY3} for $\wt \nu_\ell$.
 The key estimate  is contained in  Proposition 4.3
which depends on Proposition 4.5. The  random variables $b_j$ 
in Proposition 4.5 are now  
distributed according to $  \tilde   \nu_\ell$ and the only
 assumption of this proposition,  
the Gaussian bound (1.3) (i.e.,  the condition {\bf C1}) is 
now trivially satisfied
since $\wt \nu_\ell$ has compact support. 
 Hence we can now 
prove Proposition 4.3 using the same strategy.  Thus the equation 
for the probability estimate
appearing  after (4.6) in the paper still holds but the upper bound
on  the constant $A^2$ defined in (4.6) now 
becomes $2 M\ell^2/ (N \eta)$ due to the scaling. Thus the key estimate at
the end of the proof of Proposition 4.3 of \cite{ESY3} is now changed to 
\be
   \EE_{ \wt\nu_\ell}  \Big[ {\bf 1}_{\Omega^c}\cdot
 \P_\bb  [ |X|\ge \delta] \Big]
\leq
4\exp\big( -c\min \{ \delta\sqrt{N\eta}/\ell , \, 
\delta^2 N\eta/\ell^2\}\big)\;.
\label{lde}
\ee
Therefore, Theorem 4.1 of \cite{ESY3} holds with the estimates taking the form 
\be
\P_{ \wt \nu_\ell} \Big\{ \sup_{E \in [-(2 - \kappa)/\ell, (2 - \kappa)/\ell]}
|  \wt m^\ell_N(E+i\eta)- \wt m^\ell_{sc}(E+i\eta)| \ge \delta \Big\}
\leq C e^{-c \delta \sqrt{N\eta}/\ell}
\label{mcont-old}
\ee
for any $\delta \leq c_1 \kappa/\ell$. Passing from 
$\wt\nu_\ell$ to $  \nu_\ell$ via scaling, 
we have 
\be\label{7.1}
\P_{  \nu_\ell} \Big\{ \sup_{E \in [-2 + \kappa, 2- \kappa]}
|  m^\ell_N(E+i\eta  )-  m^\ell_{sc}(E+i\eta )| \ge \delta \Big\}
\leq C e^{-c \delta \sqrt{N\eta/\ell}}
\ee
for $\delta\le c_1\kappa$. Comparing this estimate with the original bound 
(4.1) in \cite{ESY3}, note 
 that the only change is that the $\eta$ in the exponent has deterioriated
 to $\eta/\ell$.
This is due to fact that we applied the Proposition 4.5 of \cite{ESY3} without 
taking the advantage that the variance is now reduced to $1/\ell^2$, 
which should
enhance the large deviation estimate \eqref{lde}. For our case, however,
 the estimate \eqref{7.1} is already sufficient since we are interested in the
 case $N \eta = N^{\e}$ and $\ell = (\log N)^2$. 

Finally we need to pass estimates to the original measure 
$\nu_\ell^{\otimes n}$. 
We can check that 
\[
\mbox{Var}\big(\nu_\ell^{\otimes n} , \nu^{\otimes n}\big)
 \le C n e^{- c \ell}.
\]
Since in our application $n=N^2$ and 
$\ell = (\log N)^2$, the right hand side is smaller than any negative  power of
 $N$, all necessary expectation values of observables
 w.r.t. $\nu_\ell^{\otimes n}$ can thus be passed to
 $ \nu^{\otimes n}$. This shows
 that the local semicircle 
law holds on scales $\eta \ge N^{-1+\e}$ for any $\e>0$ 
assuming only exponential bound \eqref{cond2relax} instead of the Gaussian
bound \eqref{cond2} required in {\bf C1)} of \cite{ESY3}. This
input is sufficient to conclude the proof in Section \ref{sec:timeevolved} 
if  $t=a^2$ is changed to $N^{-1+\lambda}$ in \eqref{not}.


\thebibliography{hhh}

\bibitem{BP} Ben Arous, G., P\'ech\'e, S.: Universality of local
eigenvalue statistics for some sample covariance matrices.
{\it Comm. Pure Appl. Math.} {\bf LVIII.} (2005), 1--42.

\bibitem{BI} Bleher, P.,  Its, A.: Semiclassical asymptotics of 
orthogonal polynomials, Riemann–Hilbert problem, and universality
 in the matrix model. {\it Ann. of Math.} {\bf 150} (1999): 185--266.

\bibitem{BH} Br\'ezin, E., Hikami, S.: Correlations of nearby levels induced
by a random potential. {\it Nucl. Phys. B} {\bf 479} (1996), 697--706, and
Spectral form factor in a random matrix theory. {\it Phys. Rev. E}
{\bf 55} (1997), 4067--4083.


\bibitem{D} Deift, P.: Orthogonal polynomials and
random matrices: a Riemann-Hilbert approach.
{\it Courant Lecture Notes in Mathematics} {\bf 3},
American Mathematical Society, Providence, RI, 1999

\bibitem{DKMVZ1} Deift, P., Kriecherbauer, T., McLaughlin, K.T-R,
 Venakides, S., Zhou, X.: Uniform asymptotics for polynomials 
orthogonal with respect to varying exponential weights and applications
 to universality questions in random matrix theory. 
{\it  Comm. Pure Appl. Math.} {\bf 52} (1999):1335--1425.

\bibitem{DKMVZ2} Deift, P., Kriecherbauer, T., McLaughlin, K.T-R,
 Venakides, S., Zhou, X.: Strong asymptotics of orthogonal polynomials 
with respect to exponential weights. 
{\it  Comm. Pure Appl. Math.} {\bf 52} (1999): 1491--1552.

\bibitem{Dy1} Dyson, F.J.: Statistical theory of energy levels of complex
systems, I, II, and III. {\it J. Math. Phys.} {\bf 3},
 140-156, 157-165, 166-175 (1962).

\bibitem{Dy} Dyson, F.J.: A Brownian-motion model for the eigenvalues
of a random matrix. {\it J. Math. Phys.} {\bf 3}, 1191-1198 (1962).

\bibitem{ESY1} Erd{\H o}s, L., Schlein, B., Yau, H.-T.:
Semicircle law on short scales and delocalization
of eigenvectors for Wigner random matrices.
Accepted in Ann. Probab. Preprint. {arXiv.org:0711.1730}

\bibitem{ESY2} Erd{\H o}s, L., Schlein, B., Yau, H.-T.:
Local semicircle law  and complete delocalization
for Wigner random matrices.
{\it Commun.
Math. Phys.} {\bf 287}, 641--655 (2009)


\bibitem{ESY3} Erd{\H o}s, L., Schlein, B., Yau, H.-T.:
Wegner estimate and level repulsion for Wigner random matrices.
Submitted to Int. Math. Res. Notices (2008). Preprint
{arxiv.org/abs/0811.2591}

\bibitem{ERSY}  Erd{\H o}s, L., Ramirez, J., Schlein, B., Yau, H.-T.:
{\it Universality of sine-kernel for Wigner matrices with a small Gaussian
 perturbation.} Preprint
{arxiv.org/abs/0905.2089}




\bibitem{J} Johansson, K.: Universality of the local spacing
distribution in certain ensembles of Hermitian Wigner matrices.
{\it Commun. Math. Phys.} {\bf 215} (2001), no.3. 683--705.



\bibitem{LL} Levin, E., Lubinsky, S. D.: Universality limits in the
bulk for varying measures. {\it Adv. Math.} {\bf 219} (2008),
743-779.

\bibitem{M} Mehta, M.L.: Random Matrices. Academic Press, New York, 1991.

\bibitem{PS} Pastur, L., Shcherbina, M.:
Bulk universality and related properties of Hermitian matrix models.
J. Stat. Phys. {\bf 130} (2008), no.2., 205-250.

\bibitem{Sosh} Soshnikov, A.: Universality at the edge of the spectrum in
Wigner random matrices. {\it  Commun. Math. Phys.} {\bf 207} (1999), no.3.
 697-733.

\bibitem{TV} Tao, T., Vu, V.: Random matrices: universality 
of local eigenvalue statistics. Preprint arxiv:0906.0510.


\bibitem{W} Wigner, E.: Characteristic vectors of bordered matrices 
with infinite dimensions. {\it Ann. of Math.} {\bf 62} (1955), 548-564.
\end{document}